# CRYPTANALYSIS AND ENHANCEMENT OF TWO LOW COST RFID AUTHENTICATION PROTOCOLS

Hoda Jannati and Abolfazl Falahati

Department of Electrical Engineering (DCCS Lab),
Iran University of Science and Technology, Tehran,
`{hodajannati,afalahati}@iust.ac.ir`

## ABSTRACT

*Widespread attention is recently paid upon RFID system structure considering its ease of deployment over an extensive range of applications. Due to its several advantages, many technical articles are published to improve its capabilities over specific system implementations. Recently, a lightweight anti-de-synchronization RFID authentication protocol and a lightweight binding proof protocol to guard patient safety are proposed. This contribution provides enough evidence to prove the first introduced protocol vulnerability to de-synchronization attack. It also provides the other protocol's suffering from de-synchronization attack as well as tracking the movements of the tags. This paper also addresses appropriate solutions to fix the security flaws concerning the two described protocols for secure RFID applications.*

## 1. INTRODUCTION

Today, Radio Frequency Identification (RFID) plays an important and popular role in variety of applications explicitly for its system implementation design and the ease of manufacturing. It provides identifying, target tracking, ambient condition sensing, guarding patient safety etc.

So many system-wise demandable advantages justify the means to choose the RFID system improvements as an interesting research subject. This is indeed noticeable throughout the literature on RFID [1, 2].

Furthermore, within RFID system structure context, many lightweight RFID authentication protocols [3-5] and grouping proof protocols [6-8] are proposed to be implemented over the practically required secure RFID channels. Recently, the grouping proof protocol is adopted to improve patient safety. It avoids death due to medication related errors, but such protocols suffer from de-synchronization as well as replay attacks [7, 8]. However, many security requirements such as tag synchronization together with privacy cannot be preserved by most of the published protocols for lightweight authentication scheme [9-14].

In 2010, Zhuo *et al*. proposed a lightweight anti-de-synchronization RFID authentication protocol [15]. They claimed their protocol would be secure against all attacks on RFID systems. However, in 2010, Yu *et al*. proposed a real lightweight binding proof protocol to guard the patient safety. Their protocol doesn't employ any complicated security algorithms but use only simple logic gates such as *AND* and *XOR* operations. They claimed that their proposed grouping proof protocol can resist against de-synchronization and replay attacks and can also support tag anonymity [16].

However, this contribution reveals Zhuo *et al*'s protocol vulnerability to de-synchronization attack. We show Yu *et al*'s protocol is also vulnerable to de-synchronization attack and is not secure against tracking the movements of the tags. This paper also proposes solutions to fix the security flaws just explained for both protocols.

Moreover, the next section reviews Zhuo *et al.*'s lightweight authentication protocol and its weakness, to propose appropriate solutions against the vulnerable attack. In section 3, Yuo *et al.*'s lightweight binding proof protocol with its weaknesses is described, and the proposed solutions are provided to resist the protocol from de-synchronization attack as well as tracking the movements of the tags. Finally, section 4 provides the full research work summary.

## 2. Zhuo et al.'s Lightweight Authentication Protocol (Weakness and Solution)

### 2.1. A Review of Zhuo et al.'s Lightweight Authentication Protocol

Zhuo *et al*. proposed a lightweight anti-de-synchronization RFID authentication protocol [15]. In their procedure, tags are passive and they only need to have a secure one-way hash function (i.e. a $H(.)$ function) and Exclusive-OR operations. They assume each tag (e.g., the $i^{th}$ tag) shares a secret key, $key_i$, with the backend server and each secret key is indexed by a unique index-pseudonym $IDS_i$ as well as a unique serial number $C_i$ in the associated $i^{th}$ tag backend server data base. Each tag also stores a random secret parameter $T_i$ which is only known by the $i^{th}$ tag. There are six steps in Zhuo *et al*'s lightweight authentication protocol that are briefly described below and shown in Figure 1 for justification purposes only:

Step1. A reader generates a random number $r$ to transmit it to $i^{th}$ tag.

Step2. Upon reception of $r$ by the $i^{th}$ tag, the tag computes $IDS_i = H(key_i)$, $a = H(T_i \oplus r)$ and $m = H(key_i \oplus r \oplus a \oplus C_i)$.
It then sends $IDS_i$, $a$ and the left part of the calculated massage parameter $m$, known as *m-left*, back to the reader.
At the same time, the random secret parameter $T_i$ is updated employing the right part of $m$ as $H(T_i \oplus m-right)$.

Step3. After receiving the $\{IDS_i, a, m-left\}$ by the reader, it transmit $\{r, IDS_i, a, m-left\}$ to the backend server.

Step4. Based on $IDS_i$ received, the backend server first retrieves the corresponding information of the tag, i.e. $key_i$ and $C_i$, from the local database and performs a comparison as $m-left \stackrel{?}{=} H(key_i \oplus r \oplus a \oplus C_i)-left$. If the notation is satisfied, then the backend server chooses a random number $R$ and computes $n = H(key_i \oplus R \oplus a)$.
Now, the backend server sends $R$ and *n-right* part to the reader and so computes $key'_i = key_i \oplus n-left$ and $IDS'_i = H(key'_i)$.

Finally, $key_i$ and $IDS_i$ will be updated as $key'_i$ and $IDS'_i$ respectively. The backend server stores old and new values of $\{key_i, IDS_i\}$ in its database to prevent de-synchronization problem.

Step5. This is the step that the reader has received $R$ and $n\text{-right}$ part, so it sends them to the $i^{th}$ tag.

Step6. Upon the reception of $R$ and $n\text{-right}$ part, the tag does the comparison with $n - right \stackrel{?}{=} H(key_i \oplus R \oplus a) - right$. If the notation is satisfied, then the tag computes $key'_i = key_i \oplus n - left$ and $IDS'_i = H(key'_i)$. Finally, $key_i$ and $IDS_i$ will be updated as $key'_i$ and $IDS'_i$ respectively for the next readers' call time.

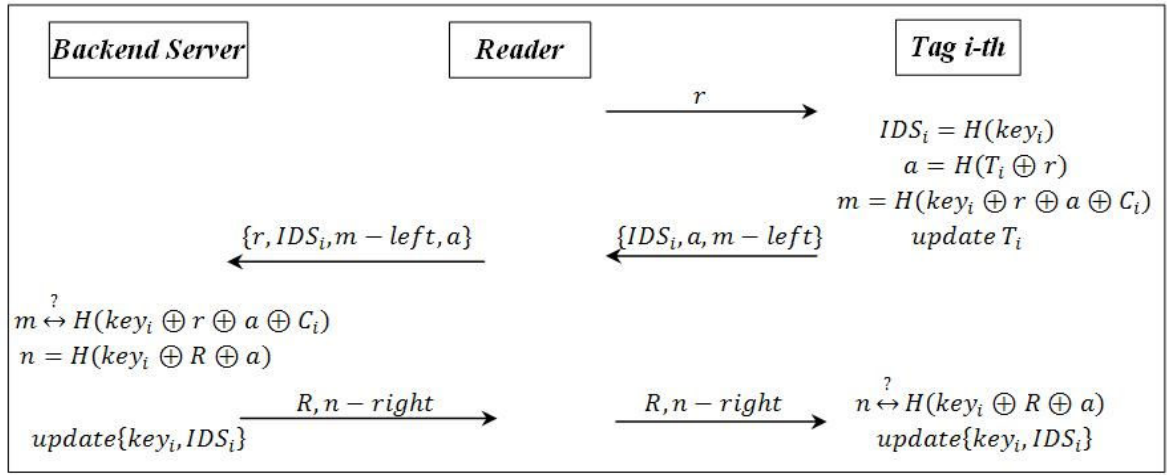

Figure 1. Communication flow for Zhuo *et al.*'s protocol [15]

## 2.2. Weakness of Zhuo et al's Lightweight Authentication Protocol

Unfortunately, the described Zhuo *et al*'s protocol is obviously vulnerable to de-synchronization attack. This can be observed by making a trivial assumption that the attacker can observe and manipulate communication link between the tags and the reader. This attack can be performed as follows;

An attacker eavesdrops $\{a, r, m\text{-left}, IDS_i\}$ by the last execution. Later, when the reader sends a random number $r'$ to the tag for the next execution, the attacker has intercepted it from the reader before forwarding it to the tag. Then, the attacker can reply to the reader instead of the tag, using the parameters obtained from the last execution (i.e. a reply attack). In this context, the attacker computes $a' = r' \oplus r \oplus a$ and then sends $a'$, $IDS_i$ and $m\text{-left}$ (in last execution) to the reader. Hence, since $a' \oplus r' = r \oplus a$, so the authentication is correctly performed which is in fact to equalize $H(key_i \oplus r \oplus a \oplus C_i) = H(key_i \oplus r' \oplus a' \oplus C_i)$. Therefore, the backend server updates $IDS_i$ and $key_i$.

Such an attack on a tag causes loss of synchronization between the tag and the backend server as shown in Figure 2. Later, when the attacked tag wants to use its key and unique index-pseudonym, the backend server identifies the tag as an illegal tag. Because, when the backend server does the comparisons $(IDS_i)_{old} \stackrel{?}{=} (IDS_i)_{new1}$ and $(IDS_i)_{new1} \stackrel{?}{=} (IDS_i)_{new2}$, undoubtedly the notations are not satisfied as it was expected. So, naturally the backend server rejects the valid tag.

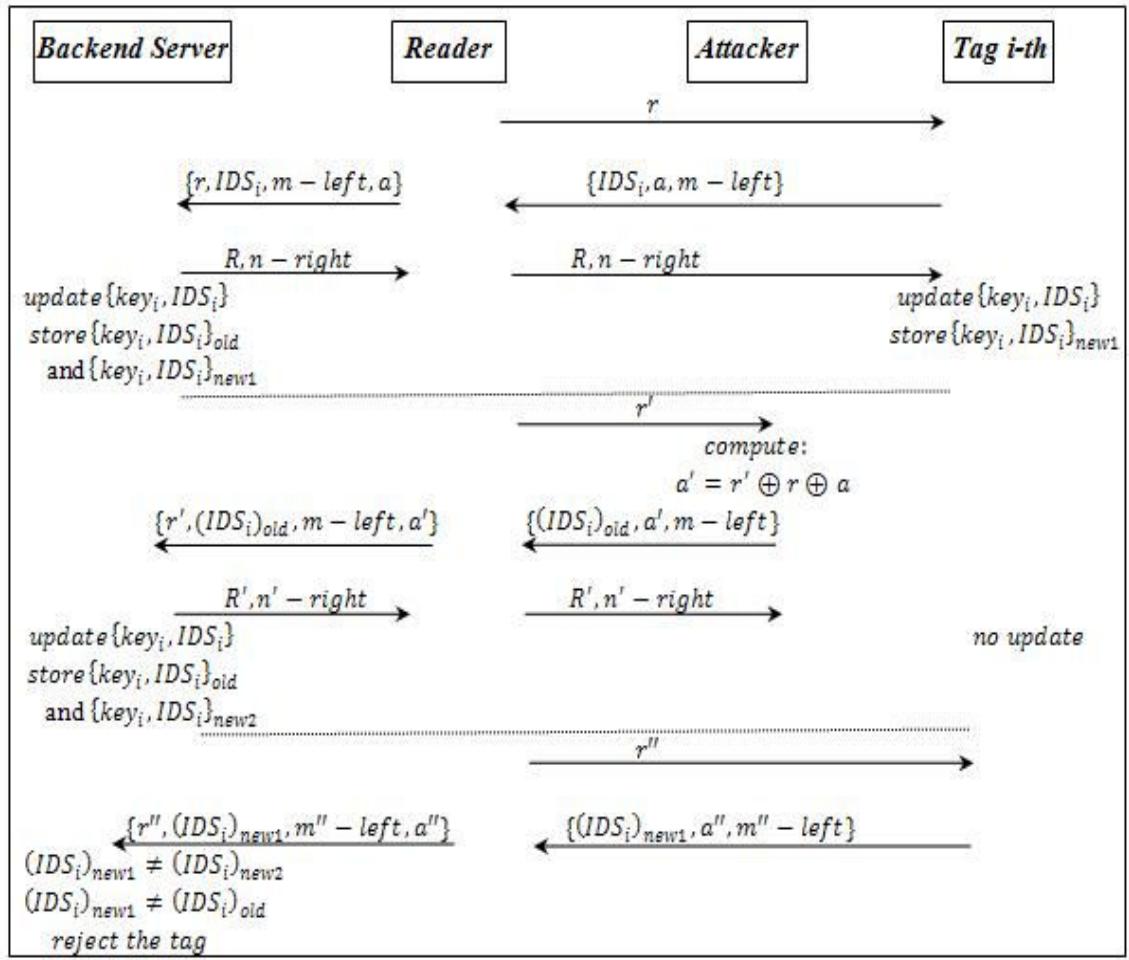

Figure 2. De-synchronization attack on Zhuo *et al.*'s protocol

In fact, the vulnerability of Zhuo *et al*'s protocol that is employed through the mentioned attack is a result of the protocol using the operation $\oplus$ in *m*. So, the attacker can change the value of *a* corresponding to *r'* sent by the reader. To overcome against the described attack, the tag must compute *m* in the form of $m = H(key_i \oplus r \oplus C_i \parallel a)$ instead. ($\parallel$ presents the concatenation operation).

Suppose that an attacker is going to use *m* again. Thus, the attacker has to modify *m* in order to show that it is generated using *r'*. However, the attacker do not know $key_i$ and $C_i$. Thus, *a'* must be found to satisfy $H(key_i \oplus r \oplus C_i \parallel a) = H(key_i \oplus r' \oplus C_i \parallel a')$. However, it is computationally infeasible since *r'* and *a'* contributes to the hash padding. So, the attacking aperture is locked up permanently.

## 3. YU ET AL.'S LIGHTWEIGHT BINDING PROOF PROTOCOL (WEAKNESSES AND SOLUTIONS)

### 3.1. A Review of Yu et al.'s Lightweight Binding Proof Protocol

Yu *et al*. proposed a lightweight binding proof protocol for medication errors and patient safety [16]. In their protocol, they assume each tag (the $i^{th}$ tag) shares $k$-bits secret keys, $X_i, k_{i1}$ and $k_{i2}$ with the backend server. They indexed the secret keys by a $k$-bits unique index-pseudonym $IDS_i$ and a $k$-bits unique identification number $ID_i$. The eight steps in Yu *et al*'s lightweight binding proof protocol are briefly described as follows and is shown in Figure 3 for justification purposes only.

Step1. The reader broadcasts "Hello" in its working range.

Step2. When the tags *A* and *B* receive the "Hello" message, both tags reply with $IDS_a$ and $IDS_b$ accordingly.

Step3. Upon the receptions of $IDS_a$ and $IDS_b$ by the reader, both parameters are then attached to the backend server to perform the operations;
$A_a = IDS_a \oplus k_{a1} \oplus r$, $B_a = ((IDS_a \vee k_{a2}) + r) \bmod 2^k$, $A_b = IDS_b \oplus k_{b1} \oplus r$ and $B_b = ((IDS_b \vee k_{b2}) + r) \bmod 2^k$ (*r* is a $k$-bits random number chosen by the backend server). At this stage, the reader transmits $\{A_a \| B_a \| IDS_b\}$ to the tag *A* and $\{A_b \| B_b \| IDS_a\}$ to the tag *B*.

Step4. Upon the reception of $\{A_a \| B_a \| IDS_b\}$ by the tag *A*, it obtains *r* from $A_a$ and then does the comparison with $B_a \stackrel{?}{=} ((IDS_a \vee k_{a2}) + r) \bmod 2^k$. If the notation is satisfied, the tag *A* computes $m_a = ((IDS_a + IDS_b + ID_a + X_a) \bmod 2^k) \oplus r$ and sends it back to the reader.

Step5. When the reader receives $m_a$, the reader sends it to the tag *B*.

Step6. When the tag *B* receives $m_a$, the tag *B* obtains *r* from $A_b$ and does the comparison with $B_b \stackrel{?}{=} ((IDS_b \vee k_{b2}) + r) \bmod 2^k$. If the notation is satisfied, the tag *B* computes $m_b = ((m_a + ID_b + X_b) \bmod 2^k) \oplus r$ and sends it back to the reader.

Step7. The reader now has $(IDS_a, IDS_b, m_a, m_b)$ to connect the tags to the backend server and prove tag *A* and tag *B* existence in the field simultaneously.

Step8. After the approval process is completed, both tags will embark into the key update process. The key update algorithm considering $(i \in \{a, b\})$ for $k_{i1}, k_{i2}$ and $IDS_i$ are computed as $k_{i1} \oplus r \oplus ((k_{i2} + ID_i) \bmod 2^k)$, $k_{i2} \oplus r \oplus ((k_{i1} + ID_i) \bmod 2^k)$ and $(IDS_i + (k_{i2} \oplus r) \oplus ID_i) \bmod 2^k$, respectively.

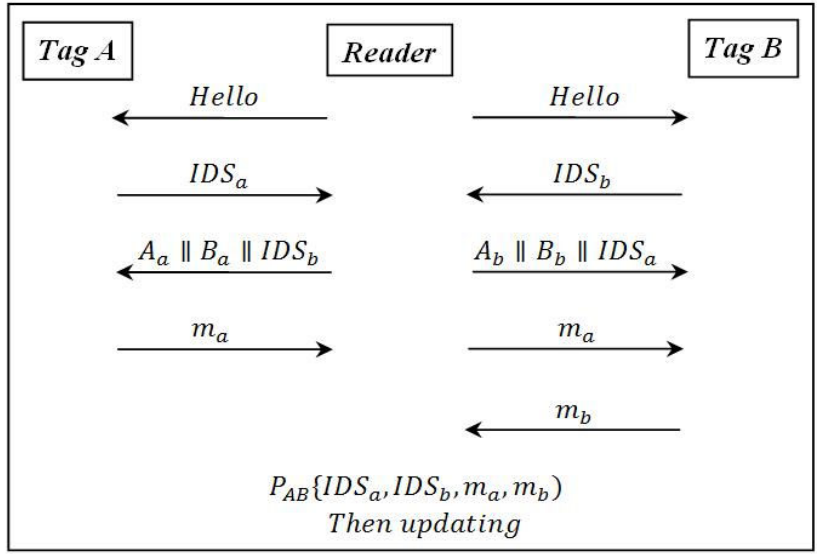

Figure 3. Communication flow for Yu *et al.*'s protocol [16]

## 3.2. Weaknesses of Yu et al.'s Lightweight Binding Proof Protocol

Unfortunately, Yu *et al*'s lightweight binding proof protocol just described is completely insecure and is susceptible to several attacks. At this stage, we introduce several attacks to Yu *et al*'s protocol to clarify our claim with the assumption that the attacker can observe and manipulate communication links between the tags and the reader.

*a. De-synchronization attack*

When the reader sends $\{A_a \| B_a \| IDS_b\}$ to the tag $A$, an attacker can replace $A_a$ and $B_a$ with $A'_a = A_a \oplus \{10000....0\}$ and $B'_a = B_a \oplus \{10000....0\}$ respectively ($\{10000....0\}$ is a $k$-bits vector with a one as the most-significant bit and all other bits with zeros).

So, the tag $A$ obtains $r' = r \oplus \{10000....0\}$ instead of $r$, because the tag $A$ computes $r$ as:

$$A'_a \oplus IDS_a \oplus k_{a1} = A_a \oplus \{10000....0\} \oplus IDS_a \oplus k_{a1}$$
$$= IDS_a \oplus k_{a1} \oplus r \oplus \{10000....0\} \oplus IDS_a \oplus k_{a1} = r \oplus \{10000....0\} = r'$$

Then the tag $A$ does comparison as $B'_a \stackrel{?}{=} ((IDS_a \vee k_{a2}) + r') \bmod 2^k$ to check the correctness of $r'$. This comparison is absolutely correct because:

$$B'_a = B_a \oplus \{10000....0\} = (B_a + 2^{k-1}) \bmod 2^k = ((IDS_a \vee k_{a2}) + r + 2^{k-1}) \bmod 2^k$$
$$= (IDS_a \vee k_{a2}) \bmod 2^k + (r + 2^{k-1}) \bmod 2^k = ((IDS_a \vee k_{a2}) \bmod 2^k + (r \oplus \{10000....0\})) \bmod 2^k$$
$$= ((IDS_a \vee k_{a2}) \bmod 2^k + (r') \bmod 2^k = ((IDS_a \vee k_{a2}) + r') \bmod 2^k$$

So, the tag $A$ obtains $r' = r \oplus \{10000....0\}$ instead of $r$ and computes $m'_a = m_a \oplus \{10000....0\}$ instead of $m_a$.

Then, when the tag $A$ sends $m'_a$ to the reader, the attacker computes $m_a = m'_a \oplus \{10000....0\}$ and sends $m_a$ back to the reader. So, the backend server uses $r$ and the tag $A$ uses $r'$ to key update processing. Therefore, during the next session, the backend server identifies the tag $A$ as an invalid tag (de-synchronization attack). This attack can be performed on the tag *B* similar to tag *A* too. These two attacks are shown in Figures 4 and 5 respectively.

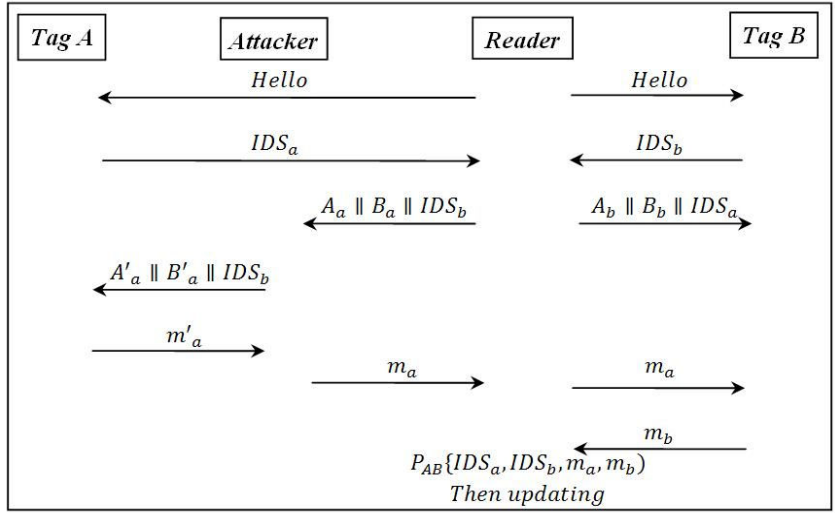

Figure 4. Tag *A* de-synchronization attack on Yu *et al.*'s protocol

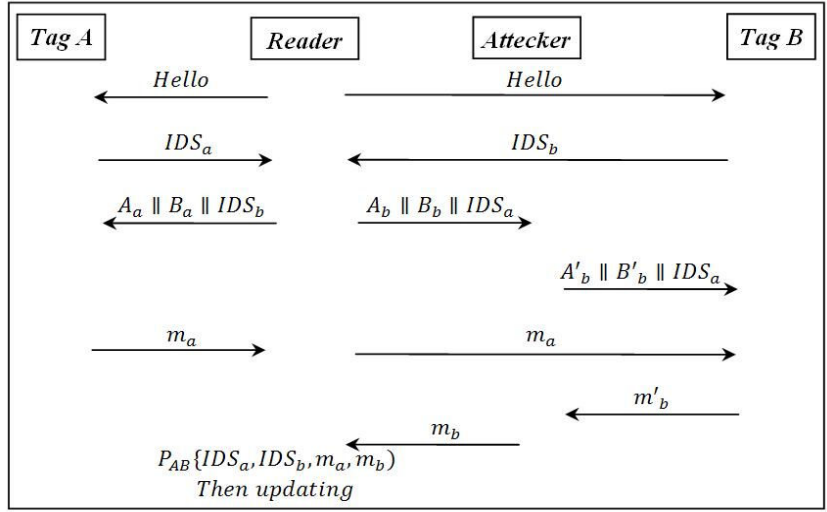

Figure 5. Tag *B* de-synchronization attack on Yu *et al.*'s protocol

b. *Tracking the movements of the tag*

Consider the attacker has an ability to track the movements of the tags *A* and *B*. For this attack, it is assumed that the given scheme is performed twice to ensure the attacker the two program executions are for the same tags or not. This technique is briefly described as follows:

It is assumed that the attacker eavesdrops the values of $(IDS_a, IDS_b, m_a, m_b)$ by two executions, e.g. executions at *i* and *j*. Consequently, the attacker will have $(IDS_a, IDS_b, m_a, m_b)_i$ and $(IDS_a, IDS_b, m_a, m_b)_j$. Then, it can compute $((m_a \oplus m_b) - IDS_a - IDS_b - m_a) \bmod 2^k$ for *i* and *j*.

If the result of $\{(m_a \oplus m_b) - IDS_a - IDS_b - m_a\}_i \bmod 2^k$ is the same as the result of $\{(m_a \oplus m_b) - IDS_a - IDS_b - m_a\}_j \bmod 2^k$, both eavesdropped parameters are executed for the same both tags, i.e. for tag *A* and tag *B*. This is for the reason that, the result of $((m_a \oplus m_b) - IDS_a - IDS_b - m_a) \bmod 2^k$ is the same as the result of $(ID_a + ID_b + X_a + X_b) \bmod 2^k$ in all executions for tag *A* and tag *B*.

Since the random number *r* has no effect on computing $A_a$, $B_a$, $m_a$ and $m_b$ and it is also only used with *XOR* operation, Yu *et al*'s lightweight binding proof protocol is vulnerable against the aforementioned attacks. Therefore, to improve Yu *et al*'s lightweight binding proof protocol to stand against described attacks, the following changes are recommended for the protocol as follows:

- *To prevent de-synchronization attack*:

    The backend server computes $A_a$ and $B_a$ in the form of $A_a = ROT((IDS_a \oplus k_{a1} \oplus r), 4) + k_{a2}$ and $B_a = ROT((IDS_a + r), 2) + k_{a2}$ and also computes $A_b$ and $B_b$ in form of $A_b = ROT((IDS_b \oplus k_{b1} \oplus r), 4) + k_{b2}$ and $B_b = ROT((IDS_b + r), 2) + k_{b2}$, respectively.

    $ROT(x, y)$ is defined to left rotate the value of *x* with *y* bits.

- *To prevent tracking the movements of the tags*:

    The tag *A* must compute $m_a$ in the form of $m_a = ((IDS_a + IDS_b + ID_a + X_a + r) \bmod 2^k) \oplus r$. Also, the tag *B* must compute $m_b$ in the form of $m_b = (((X_b \vee r) + ID_b + m_a) \bmod 2^k) \oplus r$. Since *r* is a random number, the result of $((m_a \oplus m_b) - IDS_a - IDS_b - m_a) \bmod 2^k$ is "$(ID_a + X_a + r + (X_b \vee r) + ID_b) \bmod 2^k$" and in fact this is not the same for all executions. Therefore, the attacker cannot track the tag movements.

Therefore, such modifications can lead to strengthen the protocol security against the mentioned weaknesses, i.e. de-synchronization attack and tracking the movements of the tags.

## 4. CONCLUSIONS

The investigations over two recently lightweight protocols for RFID system show that they suffer from many serious security weaknesses in spite of claims given by the protocols pioneers. These problems are found to be the two attacks called de-synchronization attack and tracking the movements of the tags. This contribution is indeed to prevent such given protocols security weaknesses employing concatenation operation on the first protocol and rotation process on the second protocol. Hence, with these improvements, both Zhou *et al*.'s and Yu *et al*.'s protocols can be used for secure RFID applications such as identifying and guarding patient safety.

**Authors**
**Hoda Jannati** received her MSc degree in Cryptography Communications from Sharif University of Technology, Tehran, Iran in 2008. Currently, she is a PhD Candidate in Electrical Engineering Department of Iran University of Science and Technology, Tehran. Her research interests include network and RFID security.

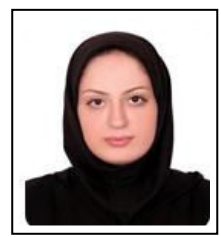

**Abolfazl Falahati** received the BSc in electronics with physics from the University of Warwick (UK) in 1980 and MSc and PhD degrees in Digital Communication Systems from Loughborough University of Technology (UK) in 1986 and 1990 respectively. He is presently an associate Professor in Cryptography, Channel Coding and Digital Signal Processing.

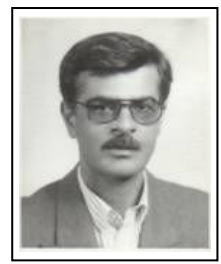